\documentclass[sigconf]{acmart}

\usepackage{comment}
\usepackage{framed}
\usepackage[framemethod=tikz]{mdframed}
\definecolor{shadecolor}{rgb}{0.75, 0.75, 0.75}
\definecolor{light-gray}{gray}{0.95}
\usepackage{float}
\usepackage{booktabs}
\usepackage{longtable}
\usepackage{soul}
\usepackage{fixltx2e}
\usepackage{flushend}
\usepackage{multibib}
\usepackage{multirow}
\usepackage{threeparttable}
\usepackage{MnSymbol,wasysym}
\usepackage{colortbl}
\usepackage{tikz}
\usepackage{pbox}
\usepackage{tabularx}
\usepackage{breakcites}

\newcolumntype{b}{X}
\newcolumntype{s}{>{\hsize=.5\hsize}X}

\newcites{sec}{Appendix. Multivocal Literature Reviews investigated}

\newcommand{\rqone}{How commonplace is to employ Grey Literature in Multivocal Literature Review studies?}
\newcommand{\rqtwo}{To what extent Grey Literature contributes with the findings of Multivocal Literature Review studies?}
\newcommand{\rqthree}{What types of Grey Literature sources are most commonly observed in Multivocal Literature Review?}

\AtBeginDocument{%
  \providecommand\BibTeX{{%
    \normalfont B\kern-0.5em{\scshape i\kern-0.25em b}\kern-0.8em\TeX}}}

\copyrightyear{2021} 
\acmYear{2021} 
\setcopyright{othergov}\acmConference[ESEM '21]{ACM / IEEE International Symposium on Empirical Software Engineering and Measurement (ESEM)}{October 11--15, 2021}{Bari, Italy}
\acmBooktitle{ACM / IEEE International Symposium on Empirical Software Engineering and Measurement (ESEM) (ESEM '21), October 11--15, 2021, Bari, Italy}
\acmPrice{15.00}
\acmDOI{10.1145/3475716.3475777}
\acmISBN{978-1-4503-8665-4/21/10}

\begin{document}






\title{What Evidence We Would Miss \\ If We Do Not Use Grey Literature?} 


\author{Fernando Kamei}
\orcid{0000-0002-5572-2049}
\affiliation{%
   \institution{UFPE, IFAL}
  \city{Macei\'o}
  \state{Alagoas}
  \country{Brazil}}
\email{fernando.kenji@ifal.edu.br}

\author{Gustavo Pinto}
\affiliation{%
  \institution{UFPA}
  \city{Bel\'em}
  \state{Par\'a}
  \country{Brazil}}
\email{gpinto@ufpa.br}

\author{Igor Wiese}
\affiliation{%
  \institution{UTFPR}
  \city{Campo Mour\~ao}
  \state{Paran\'a}
  \country{Brazil}}
\email{igor@utfpr.edu.br}

\author{M\'arcio Ribeiro}
\affiliation{%
  \institution{UFAL}
  \city{Macei\'o}
  \state{Alagoas}
  \country{Brazil}}
\email{marcio@ic.ufal.br}

\author{S\'ergio Soares}
\affiliation{%
  \institution{UFPE}
  \city{Recife}
  \state{Pernambuco}
  \country{Brazil}}
\email{scbs@cin.ufpe.br}


\renewcommand{\shortauthors}{Kamei et al.}
\renewcommand{\shorttitle}{What Evidence We Would Miss If We Do Not Use Grey Literature?}

\begin{abstract}
    \textbf{Context:} Over the last years, Grey Literature (GL) is gaining increasing attention in Secondary Studies in Software Engineering (SE). Notably, Multivocal Literature Review (MLR) studies, that search for evidence in both Traditional Literature (TL) and GL, is particularly benefiting from this raise of GL content. Despite the growing interest in MLR-based studies, the literature assessing how GL has contributed to MLR studies is still scarce.
    \textbf{Objective:} This research aims to assess how the use of GL contributed to MLR studies. By contributing, we mean, understanding to what extent GL is providing evidence that is indeed used by an MLR to answer its research question. 
    \textbf{Method:} We conducted a tertiary study to identify MLR studies published between 2017 and 2019, selecting nine MLRs studies. Using qualitative and quantitative analysis, we identified the GL used and assessed to what extent these MLRs are contributing to MLR studies.
    \textbf{Results:} Our analysis identified that 1) GL provided evidence not found in TL, 2) most of the GL sources were used to provide recommendations to solve problems, explain a topic, and classify the findings, and 3) 19 different GL types were used in the studies; these GLs were mainly produced by SE practitioners (including blog posts, slides presentations, or project descriptions).
    \textbf{Conclusions:} We evidence how GL contributed to MLR studies. We observed that if these GLs were not included in the MLR, several findings would have been omitted or weakened. We also described the challenges involved when conducting this investigation, along with potential ways to deal with them, which may help future SE researchers.
\end{abstract}



\maketitle

\section{Introduction}
    
The term ``Grey Literature'' (GL) has many definitions. The most widely accepted is the Luxembourg one~\cite{Garousi:2019:IST}, approved at the Third International Conference on Grey Literature in 1997: \textit{``[GL] is produced on all levels of government, academics, business and industry in print and electronic formats, but which is not controlled by commercial publishers, i.e., where publishing is not the primary activity of the producing body.''} The term ``grey'' (or ``fugitive'') literature is often used to refer to the literature not obtainable through traditional publishing channels, without passing through control mechanisms (e.g., peer review) before a publication~\cite{Petticrew:2006:Book:SR}. On the other hand, there are the Traditional Literature (TL), covered by peer reviewed works (e.g., conference and journal papers).

In the last years, GL gained particular attention in Software Engineering (SE) research. For instance, William and Rainer investigated the use of blogs as
an evidence source for SE research~\cite{Williams:2017:EASE, Williams:2018:ASWEC}. Several primary studies are investigating the use of GL. As an example, William and Rainer in two studies investigated the use of blogs as an evidence source for SE research~\cite{Williams:2017:EASE, Williams:2018:ASWEC}. There are also tertiary studies investigating the use of GL in secondary studies~\cite{Yasin:Thesis:2020,Zhang:2020:ICSE}.
Recently, Zhang et al.~\cite{Zhang:2020:ICSE} showed a growing number of secondary studies using GL over the years, especially the Multivocal Literature Reviews (MLR) and Grey Literature Reviews (GLR). The former is a Systematic Literature Review (SLR) that search for evidence in GL in addition to TL~\cite{Garousi:2019:IST}, while the latter only searches in GL sources. However, despite the interest, more specifically for MLR studies~\cite{Neto:2019:ESEM}, there is only one study~\cite{Garousi:2016:EASE} assessing to what extent GL sources are contributing to the findings of MLR studies.

Garousi et al.~\cite{Garousi:2016:EASE} investigated what is gained when considering GL as a source of evidence in an MLR study and what knowledge are missed when GL is not considered. However, several MLRs were published since Garousi et al.'s study, and no other research has investigated how GL affected the MLR studies. 
This lack of understanding could make SE researchers skeptical about using GL or conducting an MLR study, in particular because the addition of GL greatly increases the effort of conducting an MLR, when compared with traditional secondary studies~\cite{Raulamo-Jurvanen:2017:EASE}. 

The goal of this research is to assess to what extent the use of GL contributed to MLR studies that followed Garousi's Guidelines~\cite{garousi2017guidelines,Garousi:2019:IST}. By contributing, we mean, understanding to what extent the GL is providing evidence that is, in fact, used by an MLR to answer its research question. To achieve this goal, we explored the following research questions (RQ):

\begin{itemize}
    \item \textbf{RQ1:} \textit{\rqone} 
\end{itemize}


\begin{itemize}
    \item \textbf{RQ2:} \textit{\rqtwo} 
\end{itemize}

\begin{itemize}
    \item \textbf{RQ3:} \textit{\rqthree}
\end{itemize}

To answer these questions, we employed a tertiary study to find potential MLR studies, and qualitatively explored nine of these MLR studies. Our main findings are the following:

\begin{itemize}
    \item Several findings of MLR studies were exclusively retrieved from GL sources. For instance, we perceived that some RQs from two MLR studies~\cite{MLR4,MLR6} were answered using only GL.
    
    \item MLRs are benefiting from GL mostly to provide \emph{explanation} about a topic (e.g., explaining how DevOps could help in operations process and manage risks of companies~\cite{MLR2}) and to \emph{classify} the findings (e.g., when classifying libraries, architectural style, and architectural guidelines about Android apps~\cite{MLR3}). Also, contributions providing recommendations (e.g., a recommendation of the use of dependency injection approach to fix the heavy of the setup of test smells~\cite{MLR6}) are presented in 66.6\% of the MLR studies.
    
    \item Several GL types were identified among the MLR studies. The most common types were the blog posts, web articles, books and book chapters, and technical reports. These GLs were produced mainly by SE practitioners, consultants and companies, and tool vendors.
\end{itemize}



\section{Research Method: A Tertiary Study}\label{sec:case-studies}

As we intend to investigate to what extent GL contributed to multivocal studies, we conducted a tertiary study to identify MLR studies published in the SE literature. This research followed the most well-known guideline to conduct a secondary study in SE produced by Kitchenham et al.~\cite{Kitchenham:2007:Guideline}. 
For replication purposes, all the data used in this research is available online at: \texttt{https://bit.ly/2SBoDIh}. 

\subsection{Search strategy}\label{sec:method-search-strategy}

In this investigation, we restricted our investigation to MLR studies that strictly followed Garousi's guidelines~\cite{garousi2017guidelines,Garousi:2019:IST}. We took this decision because these are the main and most recent guidelines in SE research to conduct MLR studies. Although the most recent Garousi's guidelines were published (in a peer review format) in 2019~\cite{Garousi:2019:IST}, an earlier version of it (published in 2017 as a preprint~\cite{garousi2017guidelines}); this is why we considered both of them in our research.

We started our research in the beginning of 2020. For this reason, we decided to limit our scope to studies published since 2017 (the first publication of Garousi's Guidelines~\cite{garousi2017guidelines}) until the end of 2019.
We started by using the Google Scholar search engine to find works that cited Garousi's studies published~\cite{garousi2017guidelines,Garousi:2019:IST}. 

\subsection{Selection criteria}

When manually investigating the 60 potential studies, we focused on selecting only MLR studies.
For each candidate study, we applied a set of exclusion criteria described in Table~\ref{tab:ec}.
We excluded any candidate study that complies with at least one exclusion criterion. At the end of this process, we were left with nine MLR studies.

\begin{table}[!ht]
\caption{List of exclusion criteria.}
\small
\centering
\begin{tabular}{cp{7.5cm}}
\toprule
\textbf{\#} & Description \\
\midrule
EC1 & The study was published before 2017 or after 2019.\\
EC2 & The study was duplicated.\\
EC3 & The study was not written in English.\\
EC4 & The study was not related to Software Engineering.\\
EC5 & The study was not a full paper (e.g., a position paper).\\
EC6 & The study did not report an MLR study.\\
EC7 & The study did not follow Garousi's guidelines~\cite{garousi2017guidelines,Garousi:2019:IST}. \\
\bottomrule
\end{tabular}
\label{tab:ec}
\end{table}

\subsection{Study selection}

\begin{figure}[b!]
	\centering
	\includegraphics[scale = 0.40, clip = true, trim= 0px 0px 0px 0px]{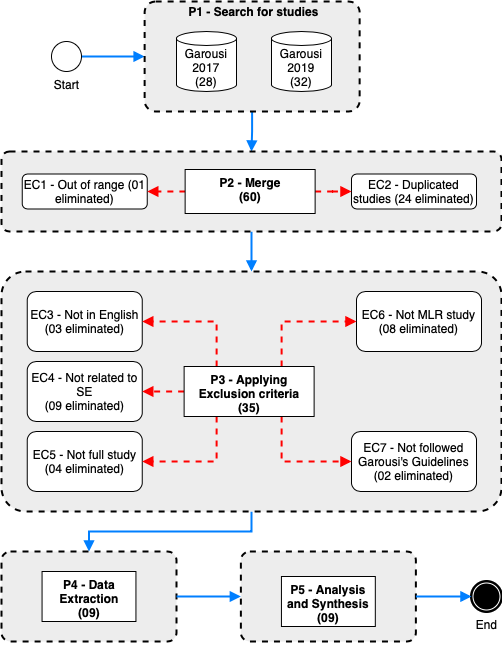}
	\caption{Process of selecting studies in each phase of the tertiary study.}
	\label{fig:selection-process}
\end{figure}

We conducted this research in five phases, as detailed in Figure~\ref{fig:selection-process}. There is a number indicating each phase (\textbf{P1--P5}).

At phase \textbf{P1}, we selected a total of 60 potential studies. From these, 28 cited the first version of the guideline for conduct MLR in SE research based on a technical report~\cite{garousi2017guidelines}, and 32 mentioned the final version of the MLR guidelines for SE~\cite{Garousi:2019:IST}.

At phase \textbf{P2}, we sorted the potential studies by title and organized them on a spreadsheet. We applied EC1 and EC2 to remove the studies out of the range of our investigation and the studies with the same bibliographical information (i.e., title, abstract, and author(s)). For EC2, we employed the following steps: (1) We compared paper titles; (2) For studies with the same title, we looked at the abstracts and if they were different. We considered the complete study as recommended by Kitchenham and Charters~\cite{Kitchenham:2007:Guideline}; if they are the same, we exclude one of them. If the publication years are different, we excluded the oldest study. We removed 25 studies, one study published after 2019 (EC1), and 24 instances of duplicated studies (EC2), respectively. At the end of this phase, 35 studies remained.

At phase \textbf{P3}, we read the studies thoroughly and applied EC3--EC7 to all the 35 potentially relevant studies. As the criteria employed to select studies were simple, only one researcher applied them alone. We removed 24 studies base on the following criteria: three studies are not written in English (EC3); nine studies are not related to SE (EC4); four studies are not full papers (EC5); six studies did not report an MLR (EC6); and two studies were eliminated because they did not follow the Garousi's studies~\cite{garousi2017guidelines,Garousi:2019:IST} to conduct their research. This way, at the end of this phase, \textbf{nine MLR studies remained}. The complete references of each study are presented in \textbf{Appendix A}.

At phases \textbf{P4--P5}, we applied the data extraction, analysis, and synthesis following the process depicted in Figure~\ref{fig:gl-delta-process}. These phases are fully described in Section~\ref{sec:delta-process}. 

\subsection{Data extraction and analysis}\label{sec:delta-process}

Due to the lack of a process to help SE researchers that intend to investigate how the use of GL contributed to MLR studies, we had to design a process based on our own experience. This process was refined by three researchers and was used to conduct phases P4 and P5. We conducted this process in pairs, and all the authors of the paper revised the emerged categories and classifications.

\begin{figure}[b]
	\centering
	\includegraphics[scale = 0.37, clip = true, trim= 0px 0px 0px 0px]{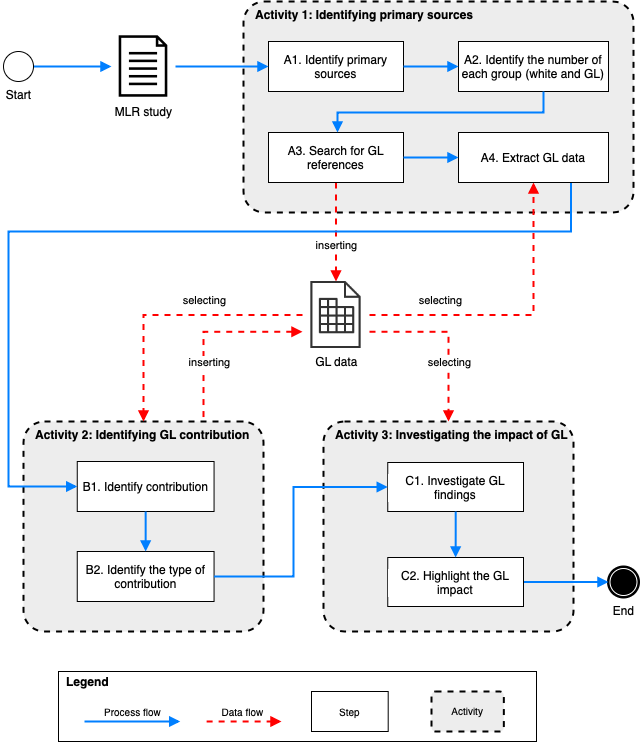}
	\caption{The process used to identify how GL use contributed to MLR studies.}
	\label{fig:gl-delta-process}
\end{figure}

Our process starts by investigating an MLR study distributed in three activities with their respective steps, as shown in Figure~\ref{fig:gl-delta-process}. In what follows, we describe our process.

\subsubsection*{Activity 1: Identifying primary sources}
The first activity aims to identify the primary sources\footnote{In related studies, the term ``primary sources'' in GL is used as equivalent to the term ``primary studies'' in TL~\cite{Garousi:2019:IST}} included in an MLR study through four steps. The first step (\textit{Step A1}) identifies the number of primary sources included in the MLR study. Then, we count the occurrences of each group: Grey Literature (GL) and Traditional Literature (TL) (\textit{Step A2}). These numbers are important in two moments: (i) to calculate the amount (\%) of GL included (total of GL included / total of included studies), and (ii) to search for GL references in the studies. The following step is to find the reference of each GL included (\textit{Step A3}), and add all the data collected in A3 in a spreadsheet. The list or references for GL is usually found in the appendix, tables, or external files available. The final step (\textit{Step A4}) consisted in selecting and extracting all the data available of each GL, in order to permit traceability between the data extracted and the primary sources (as recommended by Garousi et al.~\cite{Garousi:2019:IST}). In our research, we collected data such as (but is not limited to): (i) the names of authors,(ii) the year of publication, (iii) total number of included studies, (iv) the total number of of GL sources included, and (v) the guideline followed. In addition, considering each study that included GL, we also extracted: (i) the GL type, (ii) the evidence used from GL, (iii) the type of contribution, and (iv) type of producer.

\subsubsection*{Activity 2: Identifying the Grey Literature contribution}
The second activity consists of selecting the GL data saved to identify how its use contributed to the MLR study. Then, inserting in the spreadsheet all the portions of GL used as evidence.

We used the following approach to identify these contributions \textit{(Step B1}): (i) after identifying the GL sources, we searched for any mention/discussion of each GL in the manuscript. We noticed it is common to find this information in tables, graphics, or as citation during the manuscript; (ii) once we identified the contribution, we extracted the citation or the artifact name used to highlight where the contribution occurred; (iii) we employed a qualitative analysis to classify the contribution of the use of each GL \textit{(Step B2}) according to its type. We used the GL types classification introduced by Maro et al.~\cite{MARO:JSS:2018}; and (iv) we investigated the relation of the GL types and the contributions identified.

In the following, we present in greater detail the qualitative analysis process used in the Activity 2 (Figure~\ref{fig:code-categories}), based on the thematic analysis technique~\cite{Braun:ThematicAnalysis:2006}:

\begin{figure}[t]
\centering
\includegraphics[scale = 0.32, clip = true, trim= 90px 30px 110px 0px]{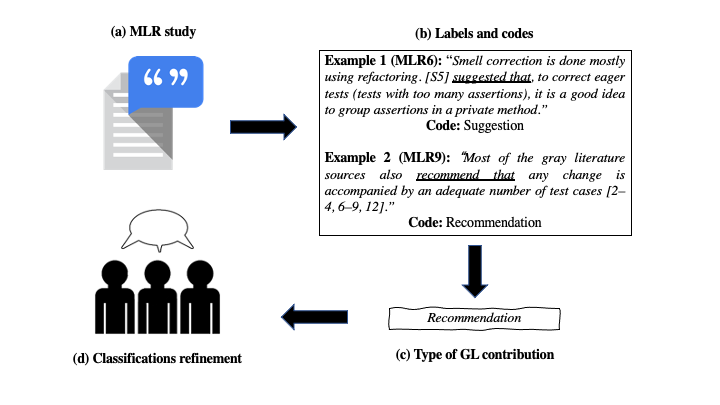}
\caption{Example of classification process used to analyze the contributions by GL use.}
\label{fig:code-categories}
\end{figure}

\begin{itemize}
    \item \textbf{Familiarizing ourselves with data.} Each researcher involved in the data analysis procedure becomes aware of which part of the MLR study the GL sources were referenced, as expressed in Figure~\ref{fig:code-categories}-(a).
   \item \textbf{Initial coding.} In this step, each researcher individually added pre-formed codes. Our process of allocating data to pre-identified themes of contributions is based on the list of contribution facets proposed by Garousi and K\"u\c{c}\"uk~\cite{GarousiBaris:JSS:2018} (e.g., recommendation, experience report, tool, solution proposal, opinion, empirical study, categorizing (or classification)). 
   During the initial coding, we found categories not identified by Garousi and K\"u\c{c}\"uk~\cite{GarousiBaris:JSS:2018}. Thus, we extended the original contribution facets to add these categories. We briefly define each one as following: 
   \textit{Programming}, used to evidence programming-related techniques; \textit{Concept Definition}, used for sources that present a concept or a definition of meaning; \textit{Explanation}, used for evidence that provides any explanation or information about a topic; \textit{Recommendation}, used for evidence that contributed by providing any recommendation to solve or support a problem or challenge. Figure~\ref{fig:code-categories}-(b) presents an example of this analysis, where two portions from the texts were extracted and coded: Suggestion and Recommendation. Labels express the meaning of excerpts from the quote that represented appropriate types of contributions.

    \item \textbf{Classifying contributions by GL use.} Here, we already had an initial list of codes. A single researcher looked for similar codes in data. Codes with similar characteristics were grouped into broader categories. Eventually, we also had to refine the categories found, comparing and re-analyzing them in parallel. Figure~\ref{fig:code-categories}-(c) presents an example of this process. This example exhibits how the category ``Recommendation'' emerged.
   
    \item \textbf{Classifications refinement.} In this step (Figure~\ref{fig:code-categories}-(d)), we involved two researchers in evaluating all classifications and a third researcher to solve any disagreements (if needed). In the cases of any doubt, we solved them through conflict resolution meetings.
\end{itemize}

\subsubsection*{Activity 3: Investigating the impact of GL}
This activity consisted of investigating how GL usage contributed to MLR study. It started by selecting the data of GL stored to investigated GL findings \textit{(Step C1)} and to understand how these findings contributed to the MLR study \textit{(Step C2)}. The goal is to assess quantitatively and qualitatively these contributions. For instance, in terms of quantitative analysis, we presented the difference in the proportion of included studies and the number of studies related to a particular finding. In qualitative aspects, we compared GL findings with TL findings, focusing on understanding if any finding was observed solely because of GL.

\section{Results}\label{sec:results}

This section answers our RQs by analyzing nine MLRs studies that followed Garousi's guidelines~\cite{garousi2017guidelines,Garousi:2019:IST}. 

First, we present an overview of how the use of GL contributed to each MLR study (Section~\ref{sec:overview-gl-contributions}). Then, we present our classification for the contributions identified and correlating them with the GL types and their producers (Section~\ref{sec:classifying-gl-contributions}). Finally, we present the types of GL and producers identified (Section~\ref{sec:gl-sources-producers}).

\begin{table}[b]
\caption{Characteristics of investigated studies. ``Total (\#)'' means the total amount of GL as the primary source, ``Total (\%)'' means the proportion of GL as primary source, ``RQ'' means the number of research questions answered with GL, and ``XRQ'' means the number of research questions \emph{exclusively} answered.}

\label{tab:mlr-studies-characteristics}
\begin{minipage}{.5\textwidth}
  \centering
  \begin{tabularx}{\textwidth}{crrss}
    \toprule
    ID & Total (\#) & Total (\%) & RQ & XRQ\\
    \midrule
    \cite{MLR1} & 15 & 32.6\% & 2/2 & 0/2\\
    \cite{MLR2} & 7 & 43.7\% & 3/3 & 0/3\\
    \cite{MLR3} & 32 & 72.7\% & 3/3 & 0/3\\
    \cite{MLR4} & 10 & 90.9\% & 3/3 & 2/3\\
    \cite{MLR5} & 120 & 72.3\% & 8/9 & 0/9\\
    \cite{MLR6} & 160 & 47.2\% & 1/3 & 1/3\\
    \cite{MLR7} & 5 & 21.7\% & 1/3 & 0/3\\
    \cite{MLR8} & 151 & 66.5\% & 1/1 & 0/1\\
    \cite{MLR9} & 21 & 48.8\% & 1/2 & 0/2\\
    \bottomrule
  \end{tabularx}
\end{minipage}
\end{table}

\subsection{RQ1: \rqone}\label{sec:overview-gl-contributions}

An overview of the nine MLR studies is presented in Table~\ref{tab:mlr-studies-characteristics}, showing several interesting observations. First, the second column (Total (\%)) shows that in the study~\cite{MLR4}, GL accounted for more than 90\% of primary sources overall. In three studies~\cite{MLR3,MLR5,MLR8}, GL accounted for between 51--75\% of the selected studies. Only one MLR study~\cite{MLR8}, GL was found in less than 25\% of included sources. This finding suggest that MLRs are taking serious advantage of GL. Second, in the third column (RQ), we depict how many GL sources were used to answer the research questions posed by the MLRs. We noticed that all studies used GL to answer at least one research question. The MLRs~\cite{MLR1,MLR2,MLR3,MLR4,MLR8}, in particular, used GL as their basis to answer all research questions. When looking closer (last column, XRQ), we also observed two studies~\cite{MLR4,MLR6} that have some RQs that were exclusively answered using GL, for instance.
Next, we assess what evidence was found in GL.

Garousi et al.~\cite{MLR1} conducted an MLR to provide a more ``holistic'' view about SE research relevance. The study included 46 primary sources, 31 from TL (67.4\%) and 15 from GL (32.6\%). Although the amount of TL studies was higher than GL sources, the evidence retrieved from GL were used to support most of the findings. The authors identified that the root causes of low relevance of SE research (e.g., Simplistic view about SE in practice, Wrong research problems identification, Issues with research mindset) were observed in multiple sources (GL and TL), concluding that the community members share similar opinions on the debate. 

Plant's study~\cite{MLR2} performed an MLR to investigate which types of risks companies using DevOps are generally exposed to and proposed a framework that helps companies control their process and manage risks. The study identified 24 risk types. From these, nine were exclusively identified in GL sources (e.g., Automated change controls and thresholds, Automate production deployment, Static code analysis), eight were exclusively identified in TL sources, and seven were found in both groups (GL and TL). In particular, if the study did not consider GL sources, the MRL would not have discussions about \textit{Automated security tests} and \textit{Monitoring and logging}, which comes largely from GL.


Verdecchia~\cite{MLR3} investigated (through an MLR and interviews with SE practitioners) how developers architect their Android apps, what architectural patterns these apps rely on, and their potential impact on quality. The study identified 15 libraries and nine architectural patterns considered when developing Android apps. Considering only the libraries, 13 were found exclusively in GL (e.g., JUnit, Mockito, Mosby), and only two of them were found through the interviews. From the architectural patterns identified, 7/9 (77.8\%) were exclusively found in GL (e.g., Clean, Hexagonal, Viper). Beyond that, 212 architectural practices were extracted and synthesized into 42 architectural guidelines. From these guidelines, 38/42 (90.5\%) were retrieved from GL. According to the study, four main themes emerge from the guidelines retrieved exclusively in GL. Regarding the quality requirements considered while architecting Android apps, seven (7/24; 29.1\%) of them were exclusively retrieved from GL (e.g., Scalability, Interoperability, Maintainability). In particualr, the scalability attribute was \emph{exclusively} found in GL sources. On the other hand, 11 groups of quality requirements were exclusively found in TL sources.

Bhandari and Colomo-Palacios~\cite{MLR4} conducted an MLR to investigate holacracy, a practice to radically shift from the conventional ladder to a more decentralized organizational structured. This MLR investigated holacracy in software development teams, its features, benefits, and challenges. This study investigated three research questions: RQ1 covered the definitions of holacracy and was answered using only GL sources. RQ2 investigated the characteristics of holacracy, which were identified: roles, circles of small groups, and meetings. Circles and meetings, in particular, were derived only from GL sources, and the roles were identified in both GL and TL. Finally, RQ3 was answered using only GL sources, explored the benefits (e.g., increased product transparency, better decisions, fast improvement) and challenges (e.g., implementation difficulty, undefined job roles cause employee outflow) using holacracy.



Garousi and K\"u\c{c}\"uk~\cite{MLR5} performed an MLR to summarize what is known about smells in test code. The authors highlighted that ``most test smells and problems in this area are `observed' by practitioners who are actively developing test scripts and are communicating by them via the GL (e.g., blog posts and industry conference talks).'' In this study, GL sources represent 72 out of 81 (88.9\%) primary sources that presented new smells names and types. For solution proposals, 72.4\% of the sources were GL.

Maro et al.~\cite{MLR6} conducted an MLR to explore traceability challenges and solutions in the automotive software development domain. The study identified 22 challenges of software traceability (e.g., Lack of knowledge and understanding of traceability, Difficulty defining information model for traceability, Unclear traceability process) distributed in seven groups of factors (e.g., Human, Uses of Traceability, Knowledge of Traceability). In this investigation, although the challenges identified in GL and TL were similar, the study mentioned that the solutions presented in GL were richer than TL due to the diversity of producers.

Freire et al.~\cite{MLR7} performed an MLR to evaluate integration platforms, specialized software tools with integration solutions, which aim to direct a set of applications to promote compatibility among their data and new features regarding the performance of their run-time systems. This study selected nine open-source integration platforms, of which two were exclusively found in GL sources (Petals and ServiceMix), five were found both GL and TL (e.g., Guaraná, Fuse, Mule), and two exclusively found in TL sources (Camel and Spring Integration). 

Saltan and Smolander~\cite{MLR8} investigated a total of 13 SaaS pricing frameworks: seven retrieved from TL (e.g., Pricing process framework, Cloud solution pricing framework) and six from GL (e.g., Customer-centric value-based pricing framework, Pricing process framework, PWC pricing management framework). These frameworks coverage the three SaaS pricing aspects (Pricing strategy, Pricing tactics, Pricing operations). Considering the pricing aspects observed, if the study did not include GL, no evidence of Pricing Operations would exist.

The study of Ram and Sawant~\cite{MLR9} focused on gaining a sound foundation about what aspects of a code change reviewers focus on, conducted two investigations: an MLR study and the other one using interviews. The study identified ten themes that constitute an excellent code change (e.g., Change description, Change scope, Code style). Two themes were identified only in TL sources (Nature of the change, Subsystem hotness). No theme was exclusively composed of GL, although in some of the themes, GL counts as the main source to provide evidence (e.g., Change description, Commit history).

\subsection{RQ2: \rqtwo}\label{sec:classifying-gl-contributions}
This section presents the results of our investigation of the 384 GL sources found in nine MLR studies, resulting in 326 contributions identified and classified. We also explored these contributions by analyzing their relation with each MLR study.

To better comprehend of the present results and enable traceability, we include direct quotes extracted from the MLR studies representing the GL use in the study. In the following, we describe each type of contribution.

\subsection*{Contributions related by GL use}

\vspace{0.2cm}
\noindent
\textbf{Recommendation (7/9 studies; 77.8\%).}
GL evidence was found by providing recommendations to deal with something (e.g., project, tool) or some problems (e.g., lack of proper visualization and reporting tools for software traceability in the automotive domain). In Garousi and K\"u\c{c}\"uk~\cite{MLR5}, the authors cited a blog post that suggested using dependency injection as an approach to fix one test smell. Maro et al.~\cite{MLR6} mentioned a service description presenting a recommendation to use a centralized data storage where all artifacts are stored and therefore accessible by the staff in different locations. This would solve the challenge of complexity added by distributed software development: \textit{``[\ldots] having tool support such as an integrated tool platform where all development activities are done, or a structured way of defining artifacts also helps to solve this challenge.''}

\vspace{0.2cm}
\noindent
\textbf{Explanation (7/9 studies; 77.8\%).} 
This category (with the highest number of contributions) indicates that authors used GL to explain some topics explored in seven MLR studies. 
An example for this category, the study of Garousi et al.~\cite{MLR1} mentioned a blog post: \textit{``Software research is biased toward huge projects and, thus, small to medium size projects may not benefit from most SE papers.''} The study of Plant~\cite{MLR2} used a whitepaper to explain how DevOps could manage risks in software companies: \textit{``[\ldots] Due to the increased speed, quality, and agility which DevOps brings about if implemented correctly, implementing DevOps processes can contribute significantly to achieving these objectives.''} In the study of Bhandari and Colomo-Palacios~\cite{MLR5}, GL sources were used to characterize holacracy in software development teams. For instance, the information present in a blog post: \textit{``In holacracy, instead of job titles, there is a strong focus on the roles that people take on within the organization. Every task or project assigned to an employee must be within the accountabilities of his or her role.''}

\vspace{0.2cm}
\noindent
\textbf{Classification (6/9 studies; 66.7\%).} 
This category was also commonly observed, indicating that GL helped to classify the findings (e.g., types of concepts, tools, SE practices) of the MLR studies. 
Verdecchia et al.~\cite{MLR3} used 32 GL primary sources and 12 TL primary studies to classify the libraries, architectural style, and architectural guidelines found about Android apps. 
As an example, the study of Verdecchia et al. used GL evidence to classify 38 architectural practices found into four themes: general Android architecture, MVP, MVVM, and Clean Architecture. In Garousi and K\"u\c{c}\"uk~\cite{MLR5}, a GL based in a bachelor thesis was used to categorize test smells, as follow: \textit{``[\ldots] categorized 53 different test smells on several dimensions, e.g., test automation, determinism, correct use of assertions, and reliability.''} 

Another example was the study of Ram et al.~\cite{MLR9} that used GL to classify the findings of what constitutes a good code change. This study used evidence from GL to classify eight themes (e.g., change description, change scope, code quality, code style).

\vspace{0.2cm}
\noindent
\textbf{Solution proposal (5/9 studies; 55.5\%).} 
In this category, the use of GL contributed to proving solutions proposals to some problems or challenges faced. 
An example for this category, the study of Maro et al.~\cite{MLR6} identified some solutions proposals for software traceability in the automotive domain, in a presentation of one company, as we quoted: \textit{``Two solutions have been suggested. One is to have tools that support the different disciplines with collaboration features such as chats, forums, and notifications. [\ldots] Second is having a defined process on how the teams should collaborate [\ldots].''} The study of Plant~\cite{MLR2} used a book to show how they implemented their DevOps process: \textit{``In order to ensure quality and information security, Muñoz and Díaz implemented phases from the OWASP Software Assurance Maturity Model (SAMM) [\ldots]. The OWASP SAMM covers the phases governance, construction, verification and operations and therefore spans the complete DevOps life cycle [\ldots].''}

\vspace{0.2cm}
\noindent
\textbf{Opinion (5/9 studies; 55.5\%).} 
This category was identified using opinions included in some GL sources. We employed the same meaning of Garousi and K\"u\c{c}\"uk~\cite{GarousiBaris:JSS:2018} for ``opinion'' contributions, in which GL sources characterizing to emit ``opinion.'' In this regard, an opinion about Android architecture based on a discussion from a blog post was used in Verdecchia et al.'s~\cite{MLR3}: \textit{``No. Do not retain the presenter I don’t like this solution mainly because I think that presenter is not something we should persist, it is not a data class, to be clear.''} Another example was presented in Garousi et al.~\cite{MLR1} that used the content of a video presentation in a conference panel as evidence. A professor in the panel emitted an opinion about the root causes of low relevance of SE research, focusing on requirements engineering in the SE area: \textit{``In my view, too often, research justified as satisfying the needs of industry begins with a wrong or simplified understanding of industry's problems.''}

\vspace{0.2cm}
\noindent
\textbf{Concept Definition (3/9 studies; 33.3\%).}
GL was used to present some concepts and definitions in MLR studies. For instance, in Bhandari and Colomo-Palacios~\cite{MLR5}, a web article presented the definition of holacracy, as followed: \textit{``The literature defined holacracy in software development teams as a way of decentralized management and organizational governance where authority and decision making are delivered throughout autonomous and self-organizing teams (circles).''} Another use of this contribution was identified in Garousi's study~\cite{MLR1}, in which a slide presentation defined the ``impact'' in SE research as \textit{``How do your actions [research] change the world?.''}

\vspace{0.2cm}
\noindent
\textbf{Experience report (3/9 studies; 33.3\%).}
To characterize the evidence found in experience-based studies, we employed the same approach of Garousi and K\"u\c{c}\"uk~\cite{GarousiBaris:JSS:2018}: \textit{``Experience studies were those who had explicitly used the term ``experience'' in their title or discussions without conducting an empirical study.''} In this regard, the study of Verdecchia et al.~\cite{MLR3} used a guideline that provided a diverse experience reports on how to test each code module (e.g., User interface and interactions, Webservice, Testing Artifacts). The study of K\"u\c{c}\"uk~\cite{MLR5} used an evidence from a blog post about unit testing that provided: \textit{``a practitioner shared her experience of moving away from assertion-centric unit testing and fixing smells such as eager tests.''}

\vspace{0.2cm}
\noindent
\textbf{Others (3/9 studies; 33.3\%).} 
Here we group the studies that the use of GL contributed with \textit{``tools''}, \textit{``code programming''}, and \textit{``empirical evidence.''} In this regard, Plant~\cite{MLR2} presented a discussion from a \textit{whitepaper} about the use of containers like Docker in DevOps, as we quoted: \textit{``They are therefore very resource efficient. However, configurations in Docker containers cannot be changed since containers cannot be updated. Updated software or configuration, therefore, requires a new image build.''} The study of Maro et al.~\cite{MLR6} used a \textit{book} that explored test smells, as following: \textit{``[GL] explored a set of ‘pitfalls’ (smells) for JUnit and an Apache-based test framework named Cactus.''} The last example is about empirical study base in a blog post, present in Garousi and K\"u\c{c}\"uk~\cite{MLR5}, in which were explored open-source projects to investigate test redundancy, as we follow: \textit{``[\ldots] [GL] reported a study on more than 50 test suites from 10 popular open-source projects and found that higher amounts of test redundancy are linked to higher amounts of bugs.''}

\subsection{RQ3: \rqthree}\label{sec:gl-sources-producers}
In our investigation, we explored: (i) the use of each GL type in MLR studies and the relation between these types with the contribution identified by GL use; and (ii) the GL types and the types of producers identified.

For a better comprehension of Table~\ref{tab:types-GL-vs-types-GL-contributions}, we informed: one GL type could be related to none, one or more of a type of contribution; and one study could be classified into none (blank), one, or in more than one type of contribution.

\subsubsection*{(i) Grey Literature vs Contributions}

\begin{figure*}[h]
    \centering
    \includegraphics[scale=0.55, clip = true, trim= 20px 0px 0px 0px]{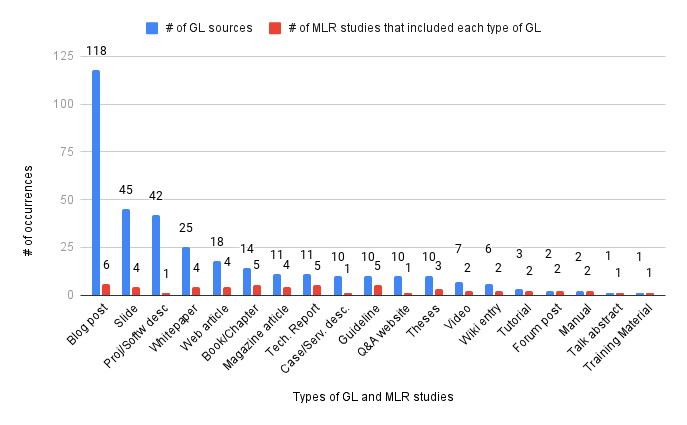}
    \caption{Amount of the Grey Literature found distributed by its types and the amount of MLR studies in which each type of GL was used.}
    \label{fig:occurences-types-gl}
\end{figure*}

We classified the 384 GL sources identified in MLR studies according to 19 types of GL. 
Figure~\ref{fig:occurences-types-gl} shows the distribution of this classification from two perspectives. The first one (blue bar) presents the amount of GL sources for each GL type. The second one (red bar) shows the amount of MLR studies in which each GL type was found. The GL types identified were related to the type of contribution identified, as shown in Table~\ref{tab:types-GL-vs-types-GL-contributions}.

Considering GL sources, \textit{Blog posts} were the most common GL type found among the MLR studies (118 occurrences), used in six MLR studies~\cite{MLR1,MLR3,MLR5,MLR6,MLR7,MLR9}. Regarding the contributions related to its use, the most commonly was to provide recommendations and opinions.

\textit{Slides presentations} was the second type most common found in the studies (45 occurrences), used in four MLR studies~\cite{MLR1,MLR5,MLR6,MLR8}. Its use was most common to provide recommendations and solution proposals.

\textit{Project or software descriptions} were the third most found type (42 occurrences), although this type was used in only one study~\cite{MLR7}. Its use provided the following contributions: solution proposals and recommendations.

\textit{Whitepapers} was another type commonly found (25 occurrences), used in four MLR studies~\cite{MLR3,MLR4,MLR6,MLR8}. The main contributions related to this use were to provide explanations, recommendations, and opinions.


\begin{table*}[]
\caption{Number of grey literature sources (separated by their types) that are related to a type of contribution.}
\label{tab:types-GL-vs-types-GL-contributions}
\begin{tabular}{lcccccccccc}
    \toprule
\multirow{2}{*}{Type of GL}                         &
\multicolumn{10}{c}{Type of contribution}                     \\
                    & REC & EXPLA & CLA & SOP & OPN & DEF & EXP & TOOLS & PRO & EMP \\
                        \midrule
Blog post & 55 & 19 & 12 & 2 & 34 & 2 & 2 & 5 & & 2 \\
Book/Chapter & 7 & 3 & 4 & 2 & 1 & & & 1 & 1 & \\
Case/Serv. desc. & 1 & & & 2 & & & & & & \\
Guideline & 17 & 5 & 5 & & 1 & & 1 & & & \\
Magazine article & & 1 & 2 & & 1 & & & & & \\
Q\&A website & 1 & 2 & & & & & & & & \\
Slide & 9 & 5 & 3 & 6 & 2 & 1 & & & & \\
Proj/Softw desc & 2 & & & 4 & & & & & & \\
Talk abstract & 1 & & & & & & & & & \\
Tech. Report & 5 & 13 & 5 & 1 & 4 & & & & & \\
Theses & & 3 & 2 & & & & 2 & & \\
Video & 4 & 8 & & & 1 & & & 1 & 3 & \\
Web article & 2 & 3 & & 3 & & 1 & & & & \\
Whitepaper & 4 & 6 & 2 & 2 & 4 & 2 & & 1 & & \\
Wiki entry & 1 & 3 & & & & & & & & \\
* Unknown & 1 & 3 & 1 & 1 & & & & & & \\
* Others & 1 & 2 & 1 & 2 & 5 & 1 & & & & \\
\bottomrule
\multicolumn{4}{l}{\begin{tabular}[c]{@{}l@{}}CLA = Classification \\ PRO = Programming \\ DEF = Concept Definition \\ EMP = Empirical Study\\ \end{tabular}} & \multicolumn{4}{l}{\begin{tabular}[c]{@{}l@{}} EXP = Experience\\ EXPLA = Explanation\\OPN = Opinion \\\end{tabular}} & \multicolumn{2}{l}{\begin{tabular}[c]{@{}l@{}}  SOP = Solution Proposal \\ REC = Recommendation\\ TOOLS = Tools\\ 
\end{tabular}}
\end{tabular}
\end{table*}

\subsubsection*{(ii) Grey Literature Producers}

\begin{figure*}
	\centering
	\includegraphics[scale = 0.55, clip = true, trim= 20px 0px 0px 0px]{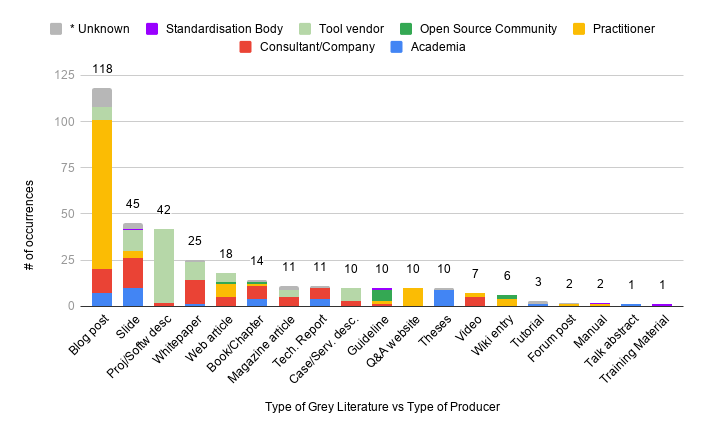}
	\caption{Distribution of each type of Grey Literature source identified among the MLR studies investigated, according to the types of producers.}
	\label{fig:types-gl-vs-type-producers}
\end{figure*}

We also investigated the producers of all 384 GL sources to identify who was the producer and to which GL types he/she was related. Figure~\ref{fig:types-gl-vs-type-producers} shows the results of these investigations.

Our first analysis shows that GL sources were produced mainly by SE \textit{Practitioners} (130/384 GL sources; 31.9\%), followed by \textit{Consultants or Companies} and \textit{Tool vendors}, each one representing respectively, 21.3\% (87/384 GL sources) and 21.1\% (86/384 GL sources).

Our second analysis showed the relationship between GL types and producer types. Three types of producers (Practitioners, Consultant or Companies, Tool vendors) caught our attention because they were responsible for almost 75\% of the GL primary sources identified. We noted that \textit{Consultants and Companies} contributed to more GL types. Their major contributions occurred with \textit{slides} and \textit{whitepapers}. \textit{Practitioners} were the second one with more contributions in different GL types. The highlighted of their contributions were mainly with \textit{blog posts}, \textit{web articles}, and \textit{Q\&A websites}. Finally, \textit{tool vendors} were the ones that most produced \textit{descriptions of projects or software} included in the MLR studies.

\section{Discussion}\label{sec:discussion}

This section revisits our main findings, discussing some of them, and relating them to related works. After, we present some challenges we faced to investigate the contributions of the GL use. Finally, we present our discussions about our limitations and threats to validity.

\subsection{Revisiting findings}

Observing the number of primary sources included in MLRs, GL has a significant contribution. Although GL sources had low inclusions rates in some studies, as in Garousi's study~\cite{MLR1}. In our opinion, it reflected the research topic investigated, which was focused on the SE research area.

We identified 19 GL types used between the MLR studies investigated. The most common types were blog posts, web articles, and book chapters, produced mainly for SE Practitioners, Consultants or Companies, and Tool vendors. These findings show that studies using GL took advantage of evidence retrieved from the practice of SE. Furthermore, GL evidence is perceived as a benefit for several studies, for including different perspectives of traditional literature and the practice of SE~\cite{Williams:2019:Journal,Zhang:2020:ICSE,Kamei:SBES:2020}. This allows reducing the bias to the propensity for only studies reporting positive findings to be published, increase reviews’ comprehensiveness and timeliness, and foster a holistic view of available evidence~\cite{Paez:2017:Medicine}.

In our investigation, our findings show that beyond the GL evidence supported some findings of TL sources, its use contributed with exclusive evidence that would not exist if GL were not investigated. It shows the importance of GL to address topics that are missing from TL sources~\cite{Sumeer:2020}. 
 
Considering the study of Verdecchia et al.~\cite{MLR3}, if they did not consider GL, no library, architectural standards, and guidelines presented on Android apps would exist since all these findings were identified only in GL and through interviews with Android SE practitioners. Moreover, in some studies~\cite{MLR4,MLR6}, GL was the only type of source that had answers to some RQs (see Table~\ref{tab:mlr-studies-characteristics}). Thus, it shows the importance of GL evidence in contributing to the synthesis of MLR studies. Although in Garousi's study~\cite{MLR1} the inclusion of GL did not add anything different from what was found in TL. For this last study, we believe this happened because all GL included were produced in \textit{Academia} by professors or researchers. 

Our investigation shows that GL has essential contributions to MLR studies by providing helpful content with recommendations, explanations, and several other contributions, although the empirical evidence is scarce. We identified that the evidence provided in MLR studies is mainly produced by SE practitioners, consultants or companies, and tool vendors. Our findings corroborate with Garousi et al.~\cite{Garousi:2016:EASE} as we found contributions related to practical solutions proposals, recommendations, opinions, or guidelines.

Previous studies investigated the types of GL sources used but not their producers. For instance, Zhang et al.~\cite{Zhang:2020:ICSE} investigated secondary studies and identified that the most common GL types used were technical reports, blog posts, books, and theses. Another tertiary study conducted by Yasin et al.~\cite{Yasin:Thesis:2020} investigated a different time-span (studies published until 2012) of our research. Our results were quite different because Yasin's study considered conference papers as a GL type. Neto et al.~\cite{Neto:2019:ESEM} also investigated MLR studies but did not investigate the amount of use for each GL type. Instead, they only mentioned that MLR studies included videos, whitepapers, books, web articles, magazine articles, blog posts, and technical reports.

\subsection{Challenges investigating Grey Literature contributions in MLR studies}

This section describes some challenges we faced to investigate the GL in MLR studies, with a possible way(s) to address each one.

The first challenge faced was to identify the GL sources included in MLR studies. This investigation in some studies was a time-consuming activity since some of them had hundreds of primary sources and some of the others did not classify the primary sources (e.g.,~\cite{MLR5}) or did not present their references (e.g.,~\cite{Saltan:IWSiB:2019,Eck:ESEC/FSE:2019}).

We recommended that SE researchers intending to conduct MLR studies to classify all primary studies/sources (TL or GL) for the first challenge. Moreover, we also recommended that GL be classified (e.g., blog post, book, theses). These recommendations are helpful for a more comprehensive understanding of GL use and to guide future researchers that may want to explore a specific GL source.

The second challenge is related to the lack of information about the GL. For instance, some essential pieces of information (e.g., the title of the source, URL, last accessed, name of the author(s), type of GL, type of producer) were not available for several GL sources in MLRs studies~\cite{MLR8,Saltan:IWSiB:2019,Eck:ESEC/FSE:2019}. This challenge precludes a better understanding of each GL source and answers our research questions. For this reason, we removed these studies~\cite{Saltan:IWSiB:2019,Eck:ESEC/FSE:2019} from our analysis, although they presented some important information about GL in their studies. For instance, Saltan~\cite{Saltan:IWSiB:2019} investigates challenges about flaky tests, mentioning the high number of relevant GL sources identified compared with TL sources, which shows that flaky test understanding is still scarce.

To address the second challenge, we recommended to the researchers include all the information available from GL sources. This information may be essential for the reader to better understand the GL source used and guide future research to a deep investigation of GL sources.

The third challenge relates to identifying and classifying contributions by GL use, which is a consequence of the first two challenges. For instance, it was not possible to conduct a deep investigation of the GL sources in two MLR studies~\cite{Saltan:IWSiB:2019,Eck:ESEC/FSE:2019}. Moreover, we perceived that the studies often did not highlight the differences between the findings from GL and TL.

One possible way to address the third challenge is following the Garousi et al.'s guidelines~\cite{Garousi:2019:IST} which recommended that the data extraction be conducted separated by the different types of source (GL and TL) and a balanced synthesis using sources with varying levels of rigor. In our opinion, another possibility is the synthesis highlight the differences between GL and TL, aiming to the reader understand how each type of primary source contributed to the study and the relevance of each piece of evidence presented.

\subsection{Limitations}
This section discusses the potential threats to the validity of our study and what we have done to minimize or mitigate them. 

An internal threat of any qualitative investigation is related to the interpretation. The case of our research relates to how we interpret the contributions identified by GL use. As this activity involves personal understanding, to mitigate this threat, we followed a paired process during this research, and a third researcher revised the derived categories.

An external threat is related to the impossibility of precisely determining all GL use contributions because, in several MLR studies, both GL and TL were not referenced in the articles. We know that it is common to occur in any secondary studies, mainly in that study with several studies included. In some studies (e.g.,~\cite{Eck:ESEC/FSE:2019}) the list of primary sources was not available. We tried to mitigate this threat by sending mail to the studies. Another threat is related to our decision to select only MLR studies that followed Garousi's guidelines to investigate studies that followed a well-known process to conduct a multivocal review in SE. This decision might have introduced a bias in our findings, limiting the discussions' scope about the contribution and types of GL identified.

\section{Related Works}\label{sec:related_works}

GL investigations in SE research are particularly recent~\cite{Garousi:2016:EASE,Williams:2018:EASE}. In the context of studies that investigated MLRs studies, we found three studies~\cite{Kitchenham:2009:ESEM,Garousi:2016:EASE,Neto:2019:ESEM} that are more related to this research.

Kitchenham et al.~\cite{Kitchenham:2009:ESEM} conducted one of the first studies using the multivocal approach in SE, comparing the use of manual and automated searches and assessing the importance and breadth of GL. Their findings showed the importance of GL, especially to investigate research questions that need practical and technical answers. For instance, when comparing two technologies. Although they recognized that, in general, the quality of GL studies is lower than TL.

Garousi et al.~\cite{Garousi:2016:EASE} expanded the investigation of GL as a source of evidence for MLR studies in SE research, conducting two investigations. The first one presented a meta-analysis from three cases in which GL was used to understand what knowledge is missed when an SLR does not consider GL. The second one investigated three MLRs to understand what the community gains when conducting multivocal literature. The study highlighted the importance of using GL to cover technical research questions and be beneficial to practitioners, once the evidence is retrieved from the industry.

Neto and colleagues~\cite{Neto:2019:ESEM} investigated MLRs and GLRs studies through a tertiary study. Their research aimed to understand (i) the motivations to included GL (lack of academic research on the topic, practical evidence, emerging research on the topic), (ii) the types of GL used (videos, tools overview, blog posts, books, industrial journals, technical reports, and websites), and (iii) the search engines used, mainly focused on Google's regular search engine. They searched for the studies published between 2009 and 2019 using six academic search engines. From 56 studies returned, they selected 12.

Other studies were conducted to investigated GL in secondary studies, in general, not specifically focused on MLR studies~\cite{Yasin:Thesis:2020,Zhang:2020:ICSE}. The first one was conducted by Yasin et al.~\cite{Yasin:Thesis:2020} investigated the extent of GL use in secondary studies published until 2012 and the importance of Google Scholar to identify GL sources. In the period analyzed, the perceptions of GL in SE research and its types were different to nowadays. For instance, Yasin et al. considered workshop papers and conference proceedings as GL. These types and the technical reports were the common types identified in the investigated studies. Google Scholar was not considered a vital source to identify these sources. The second one was conducted by Zhang et al.~\cite{Zhang:2020:ICSE} that investigated GL through a tertiary study and survey research, focusing on understanding the GL definitions used in the studies and the types of GL used. The study did not identify a standard definition, and the most common GL types identified were technical reports, blog posts, books, and theses.

In SE research, few works investigated MLR studies to compare the contributions perceived by the use of GL and TL. For this reason, this research intends to expand and improve the knowledge in this regard, adding investigations and new explored topics. Our research differs from the previous ones by: (i) investigating all GL evidence included to understand and classify their contributions in MLR studies; and (ii) providing a process to support SE researchers that intend to investigate the contributions of GL in secondary studies.

\section{Conclusions and Future Work}\label{sec:conclusions}

In this paper, we conducted a tertiary study with MLR studies to better understand the GL and its contributions to these studies. We investigated a total of nine MLR studies that followed Garousi's guidelines.

Our analysis consisted of comparing the findings from GL and TL and analyzing and classifying their contributions in with each study. Our results are important to comprehend the impacts of GL use in MLR studies and increase the state-of-art by pilling additional evidence on this topic.

Our findings show that GL use stood out as an essential source to contribute with recommendations, explanations, solutions proposals, and opinions about a topic. Beyond permitting the state of the practice to be included in MLR studies, once most of the GL sources investigated were produced by SE practitioners, consultants or companies, and tool vendors. 

We identified that several of these contributions were exclusively found in GL sources. Thus, if the studies did not consider GL, several findings would not have been introduced, making the results potential biased. Moreover, GL also supported several findings found in the TL.

This study has some implications for SE research. First, by describing the process used and the challenges we faced to investigate the GL usage contributions to MLR studies, we hope to help SE researchers to take advantage of this type of investigation. Moreover, we provided additional evidence to show how GL use contributed to MLR studies.

For future works, our intention includes expanding our view to MLR studies that did not follow Garousi's guidelines~\cite{garousi2017guidelines,Garousi:2019:IST}, to investigate MLR studies authors to understand their perceptions about GL use.

\bibliographystyle{ACM-Reference-Format}
\bibliography{references}

\appendix


\nocitesec{*}
\bibliographystylesec{appendixStyle}
\bibliographysec{secondaryStudies}

\end{document}